\documentclass[fleqn,12pt]{article}
\usepackage{epsfig}
\begin{document}
\textheight 22cm
\textwidth 15cm
\noindent
{\Large \bf The role of magnetic shear for zonal flow generation}
\newline
\newline
J. Anderson\footnote{anderson.johan@gmail.com}, H. Nordman$^{\dagger}$
\newline
Department of Fundamental Energy Science
\newline
Graduate School of Energy Science, Kyoto University, Gokasho, Uji, Kyoto 611-0011
\newline
$^{\dagger}$ Department of Radio and Space Science, EURATOM-VR Association
\newline
Chalmers University of Technology, G\"{o}teborg, Sweden
\newline
\newline
\begin{abstract}
\noindent
The role of magnetic shear for zonal flow generation by ion-temperature-gradient (ITG-) and trapped electron (TE-) mode turbulence is studied analytically using fluid descriptions. The scaling of the zonal flow (ZF) growth rate with magnetic shear is examined and compared with linear growth rates for typical tokamak parameter values. The results indicate that large levels of ZF are obtained in regions of negative magnetic shear, in particular for ZF driven by TE mode turbulence. The strong magnetic shear scaling obtained for TE mode driven zonal flows originates from the bounce average of the electron magnetic drifts.
\end{abstract}
\newpage
\section{Introduction}
The study of turbulence suppression and generation in magnetically confined fusion plasma research has attracted much interest recently. It is well established that the interaction between small scale turbulence and zonal flows, which are radially localized flows propagating mainly in the poloidal direction, can reduce the radial transport by shearing the eddies of the driving background turbulence~\cite{a11}-~\cite{a131}. Accordingly, the self-consistent generation of zonal flows by non-linear interactions among drift waves has recently been studied both analytically~\cite{a14}-~\cite{a23} and numerically ~\cite{a24}-~\cite{a32}.

The purpose of the present letter is extend previous analytical studies~\cite{a19}-~\cite{a191} to investigate the effects of magnetic shear on the zonal flow generation by ion-temperature-gradient (ITG-) and trapped electron (TE-) mode turbulence, respectively. The ITG- and TE-mode driven cases will be considered separately and the corresponding ZF generation by ITG and TE mode turbulence will be calculated and compared.

The fluid models used in the present Letter have been extensively described in previous papers and are only briefly summarized here. For the ITG mode description, a reduced version of the standard Weiland model~\cite{a35} is utilized. In this formulation the ITG model includes the ion continuity and the ion temperature equations, neglecting the influence of parallel ion momentum and electromagnetic effects, while the electrons are assumed to be Boltzmann distributed. The TEM model is based on the bounce averaged electron continuity and electron temperature equations for the trapped electron fluid~\cite{a35}. The ion and trapped electron mode fluid models are fairly symmetric except for the FLR terms which are included in the ion fluid model and the bounce average which is performed on the electron quantities. The equations are:
\begin{eqnarray}
\frac{\partial n_{j}}{\partial t} +\nabla \cdot \left( n_{j} \vec{v}_{E} + n_{j} \vec{v}_{\star j} \right)+ \nabla \cdot  \left( n_j \vec{v}_{Pj} + n_j \vec{v}_{\pi j}\right)  & = & 0 \\
\frac{3}{2} n_{j} \frac{d T_{j}}{dt} + n_{j} T_{j} \nabla \cdot \vec{v}_{j} + \nabla \cdot \vec{q}_{j} & = & 0 \\
q_j =  \frac{5}{2} \frac{p_j}{m_j \Omega_{cj}} \left( e_{\parallel} \times \nabla T_j\right)
\end{eqnarray}
where $n_j$, $T_j$ are the density and temperature perturbations ($j=i$ and $j=et$ represents ions and trapped electrons) and $\vec{v}_j = \vec{v}_E + \vec{v}_{\star} + \vec{v}_{Pj} + \vec{v}_{\pi j}$, $\vec{v}_E$ is the $\vec{E} \times \vec{B}$ velocity, $\vec{v}_{\star}$ is the diamagnetic drift velocity, $\vec{v}_{Pj}$ is the polarization drift velocity, $\vec{v}_{\pi j}$ is the stress tensor drift velocity and $\vec{q}_j$ is the heat flux. The derivative is defined as $d/dt = \partial / \partial t + \rho_s c_s \vec{z} \times \nabla \tilde{\phi} \cdot \nabla$ and $\phi$ is the electrostatic potential.
The linear solutions to Eqs 1 - 3 for TE modes and ITG modes respectively are given by
\begin{eqnarray}
\omega_{r TEM} & = & - \frac{k_y}{2} \left( \xi \left(1 - \epsilon_n g_e\right) - \frac{10}{3} \epsilon_n g_e \right)  \\
\gamma_{TEM} & = & k_y \sqrt{\xi \epsilon_n g_e \left( \eta_e -\eta_{e th} \right )} \\
\eta_{e th} & = & \frac{2}{3} - \frac{\xi}{2} + \frac{10}{9 \xi}\epsilon_n g_e
+ \frac{\xi \epsilon_n g_e}{4} + \frac{\xi}{4 \epsilon_n g_e} \\
\omega_{r ITG} & = & \frac{k_y}{2\left( 1 + k_{\perp}^2\right)} \left( 1 - \left(1 + \frac{10}{3\tau} \right) \epsilon_n g_i - k_{\perp}^2 \left( \alpha_i + \frac{5}{3\tau} \epsilon_n g_i \right)\right)  \\
\gamma_{ITG} & = & \frac{k_y}{1 + k_{\perp}^2} \sqrt{\frac{\epsilon_n g_i}{\tau} \left( \eta_i - \eta_{i th}\right)} \\
\eta_{i th} & \approx & \frac{2}{3} - \frac{\tau}{2} + \frac{\tau}{4 \epsilon_n g_i} + \epsilon_n g_i \left( \frac{\tau}{4} + \frac{10\tau}{9}\right)
\end{eqnarray}
where $\omega = \omega_r + i \gamma$. Here, $\epsilon_n = L_n/R$, $\eta_j = L_n/L_{Tj}$, $\tau = T_e/T_i$,  $\alpha_i = \frac{1}{\tau} \left( 1 + \eta_i \right)$, $\xi$ is the fraction of trapped electrons and $g_j$ ($j = i,e$) are defined below.
 
The zonal flow evolution is treated by the vorticity equation. The generation of ZF by ITG- and TE-modes are studied separately for $\eta_i >> \eta_e$ (ITG dominated case) and $\eta_e >> \eta_i$ (TEM dominated case) respectively. In these limits it is assumed that the electron and ion responses are decoupled and hence the effects of ion perturbations on the TE mode are neglected (and vice versa). The previous work on zonal flow generation by ITG and TE modes~\cite{a19}-~\cite{a191} is here extended to include the magnetic shear dependence through the magnetic drift frequencies and the perpendicular wave vector. The effect of magnetic shear enters through the averaged magnetic drift frequency for the electrons and ions. The ion fluid model is averaged using an approximate eigen-mode function resulting in an averaged magnetic drift frequency: $\psi = \frac{1}{\sqrt{3 \pi}}(1+\cos(\theta))$ and $|\theta|< \pi$, giving $\langle g_{i}\rangle \propto \epsilon_n (\frac{2}{3}+\frac{5}{9}s - \epsilon)$ and $\langle k^2_{\perp}\rangle \propto (1 - 2.5 s^2+\frac{\pi^2}{3}s^2)$~\cite{a45}. Here $s$ is the magnetic shear and $\epsilon = a/R$ is the inverse aspect ratio parameter. The effects on zonal flow generation of magnetic shear have been treated for ITG modes in Ref.~\cite{a22} using the reductive perturbation method for the zonal flow generation. 

The electron fluid equations are bounce averaged and the electron magnetic drift is replaced by the precession frequency of trapped electrons $\langle \omega_{De} \rangle = \omega_{De 0} g_e$  where $\omega_{De 0} = \frac{2 k_{\theta} \rho_s c_s}{R}$ and $g_e = \frac{1}{4} + \frac{2}{3}s$~\cite{a46}. The original formulation of the Weiland model for deeply trapped electrons is recovered for $s = 9/8$ in this model. 

The zonal flow dispersion relations are found using the vorticity equation and combining this with the wave kinetic equation as an additional relation between the background turbulence and the zonal flow. The average of the ion continuity equation over the magnetic surface and over fast small scales employing the quasi-neutrality the evolution of the mean flow is obtained
\begin{eqnarray}
\frac{\partial}{\partial t} \nabla_x^2 \Phi -\mu \nabla_x^4 \Phi = \left(1 + \frac{\delta_{im}}{\tau} \right) \nabla_x^2 \left<\frac{\partial}{\partial x} \tilde{\phi}_k \frac{\partial}{\partial y}\tilde{\phi}_k \right> + \frac{\delta_{im}}{\tau} \nabla_x^2 \left<\frac{\partial}{\partial x} \tilde{\phi}_k \frac{\partial}{\partial y}\tilde{T}_{ik} \right>
\end{eqnarray}
where it is assumed that only the small scale self interactions are the important interactions in the RHS. Here $\delta_{im}$ is the Kronecker delta with $m = \{ i, et\}$. It is also assumed that there is a sufficient spectral gap between zonal flows and the background turbulence. In the case of zonal flows ($\Omega$) generated from ITG mode turbulence the dispersion relation is~\cite{a19}
\begin{eqnarray}
\left(\Omega + i \mu q_x^2 \right)\left(\Omega - q_x v_{gx}\right)^2 = - q_x^2 \left(1 + \frac{1}{\tau}  + \frac{1}{\tau} \delta \right) |\tilde{\phi}_k|^2 \Omega 
\end{eqnarray}
and in the TEM generation case we obtain~\cite{a191}
\begin{eqnarray}
\Omega = i q_x k_y\sqrt{|\tilde{\phi}_k|^2}.
\end{eqnarray}
Here, $\Omega$ is the zonal flow growth rate, $k_x, k_y$ is the wave numbers for the background turbulence, $q_x$ is the wavenumber of the zonal flow, $v_{gx}$ is the group velocity for the drift wave and $\mu$ is a corresponding collisional damping (which is neglected in this study). The $\delta $ is a wave number independent factor given by
\begin{eqnarray}
\delta & = & \frac{\Delta_k k_y}{\Delta_k^2 + \gamma_k^2} \left(\eta_i - \frac{2}{3} \left(1 + \frac{1}{\tau} \right)  \epsilon_n g_i\right) \\
\Delta_k & = & \frac{k_y}{2}\left( 1 - \epsilon_n g_i + \frac{4}{3\tau} \epsilon_n g_i\right).
\end{eqnarray}
Hence, the zonal flow growth rate scales as $\Omega \propto |\phi_k|$. Proceeding from here, several different assumptions are possible on the saturation level for the background turbulence in absence of zonal flows. In the present work only the mixing length saturation level ($\tilde{T}_j = \frac{1}{k_x L_{T_j}}$) is considered.
\section{Results}
The dispersion relations (Eqs 1 and 2) for the zonal flow ($\Omega$) is solved numerically. Special attention is given to the magnetic shear dependency of the zonal flows generated from ITG mode and TEM turbulence, respectively. 

In Figure 1, the magnetic shear scaling of the zonal flow growth rate driven by ITG modes (normalized to the linear ITG growth rate) is displayed. The parameters are $k_x = k_y = 0.3$, $q_x=0.3$, $\tau = 1.0$ and $\epsilon = 0$. In Figure 1a the other parameters are $\eta_e = 0$ (suppressing the TE mode), $\eta_i = 4.0$, $f_t = 0$ (fraction of trapped electrons ) and $\epsilon_n$ is given by $\epsilon_n = 1.5$ (rings), $\epsilon_n = 1.0$ (asterisk) and $\epsilon_n = 0.5$ (diamond).

\begin{figure}
  \includegraphics[height=.3\textheight]{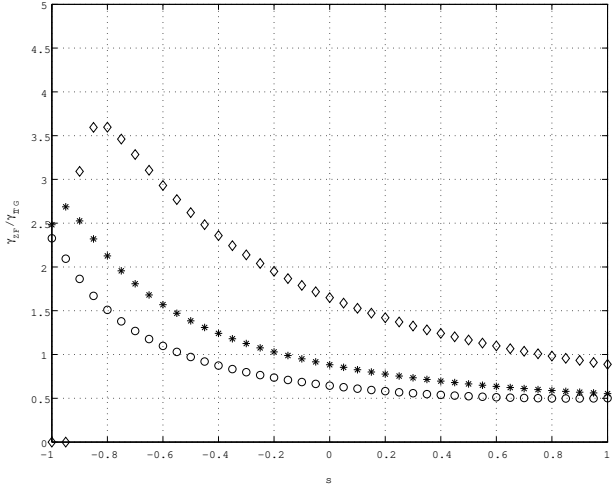}
  \caption{The ZF growth (normalized to the ITG mode growth) as a function of magnetic shear $s$ with $\epsilon_n$ as a parameter. The results are shown for $\epsilon_n = 1.5$ (rings), $\epsilon_n = 1.0$ (asterisk) and $\epsilon_n = 0.5$ (diamonds). The other parameters are $k_x = k_y = 0.3$, $q_x=0.3$, $\epsilon = 0$, $\eta_e = 0$, $\eta_i = 4.0$ and fraction of trapped electrons $f_t = 0$.}
\end{figure}
 
The corresponding result for the TEM driven case is shown in Figure 2 for $\eta_i = 0$ (suppressing the ITG mode), $\eta_e = 4.0$ and $f_t = 0.5$. The parameter $\epsilon_n$ is given by $\epsilon_n = 2.0$ (plus), $\epsilon_n = 1.5$ (rings) and $\epsilon_n = 1.0$ (asterisk). For sufficiently large negative shear the linear TEM is stabilized due to the change of sign of the precession frequency. For the zonal flows generated by TEM background turbulence there is a significant increase in growth rate in the weak and negative shear regions where $\gamma_{ZF}>>\gamma_{TEM}$. For zonal flows generated by ITG turbulence on the other hand, only a modest increase in the corresponding region is found. For positive magnetic shear, the ZF scaling with magnetic shear is rather weak. In general we obtain $\gamma_{ZF}(ITGM)>\gamma_{ZF}(TEM)$ in this region. We have verified that the ZF growth scales weakly with the other parameters. 

\begin{figure}
  \includegraphics[height=.3\textheight]{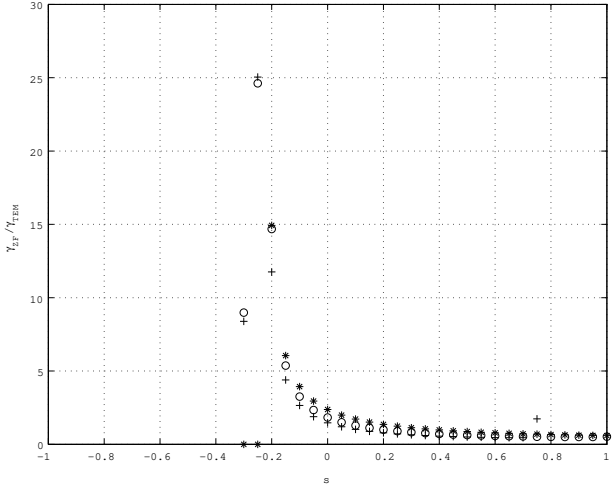}
  \caption{The ZF growth (normalized to the TEM growth) as a function of magnetic shear $s$ with $\epsilon_n$ as a parameter. The results are displayed for $\epsilon_n = 2.0$ (plus), $\epsilon_n = 1.5$ (rings) and $\epsilon_n = 1.0$ (asterisk). The other parameters are $k_x = k_y = 0.3$, $q_x=0.3$, $\epsilon = 0$, $\eta_i = 0$, $\eta_e = 4.0$ and fraction of trapped electrons $f_t = 0.5$.}
\end{figure} 

Next, the scaling of zonal flow generation with magnetic shear with normalized temperature gradient as a parameter is displayed. The parameters are $k_x = k_y = 0.3$, $q_x=0.3$, $\tau = 1.0$ and $\epsilon = 0$. In Figure 3 the other parameters are $\eta_e = 0$, $\epsilon_n = 1.0$, $f_t = 0$ with $\eta_i = 6.0$ (squares) and $\eta_i = 4.0$ (asterisk). The corresponding result for the TEM driven case is displayed in Figure 4 for $\eta_i = 0$, $\epsilon_n = 1.0$ and $f_t = 0.5$. The parameter $\eta_e$ is given by $\eta_e = 6.0$ (squares) and $\eta_e = 4.0$ (asterisk). As found in Figures 1a and 1b for weak/negative magnetic shear, a substantial increase in the zonal flow generation for TEM turbulence whereas a modest increase is found for ZF driven by ITG turbulence. The dependence on normalised temperature gradient is weak except close to the linear threshold.

\begin{figure}
  \includegraphics[height=.3\textheight]{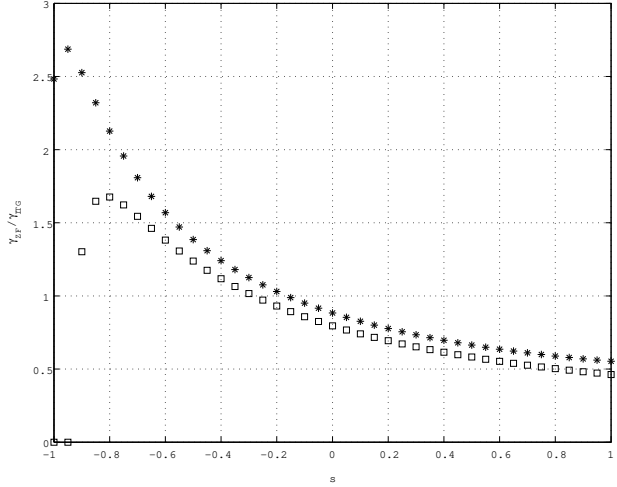}
  \caption{The ZF growth (normalized to the ITG mode growth) as a function of magnetic shear $s$ with normalized temperature gradient as a parameter. The results are shown for $\eta_i = 6.0$ (squares) and $\eta_i = 4.0$ (asterisk).  The other parameters are $k_x = k_y = 0.3$, $q_x=0.3$, $\epsilon = 0$, $\eta_e = 0$, $\epsilon_n = 1.0$ and fraction of trapped electrons $f_t = 0$.}
\end{figure}

\begin{figure}
  \includegraphics[height=.3\textheight]{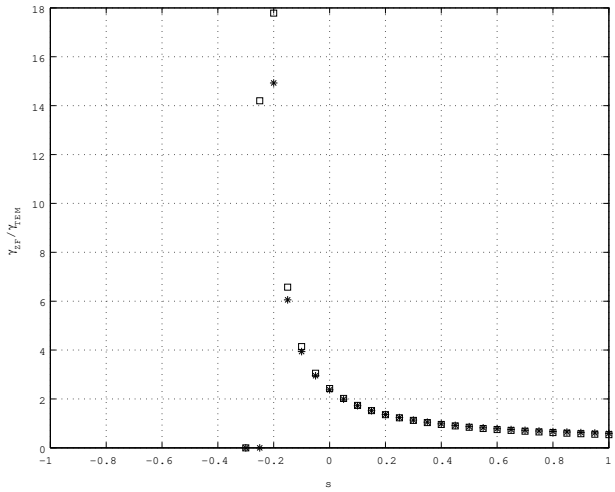}
  \caption{The ZF growth (normalized to the TEM growth) as a function of magnetic shear $s$ with normalized temperature gradient as a parameter. The results are shown for $\eta_e = 6.0$ (squares) and $\eta_e = 4.0$ (asterisk).  The other parameters are $k_x = k_y = 0.3$, $q_x=0.3$, $\epsilon = 0$, $\eta_i = 0$, $\epsilon_n = 1.0$ and fraction of trapped electrons $f_t = 0.5$.}
\end{figure}

\section{Summary}
In summary, the results indicate that the effects of magnetic shear for ZF generation are significantly different for ITG- and TE-mode driven cases. This difference is related to the different shear dependence of the ion and electron magnetic drift frequency respectively.  Hence, for the TE mode dominated case, a weak/negative magnetic shear results in a large zonal flow growth rate ($\gamma_{ZF} > \gamma_{TEM}$). This may result in a strong stabilization of the turbulence. For the ITG dominated case on the other hand, the results indicate that the effects of magnetic shear on the zonal flow growth rate is rather weak. Hence, additional effects of negative magnetic shear, not included in the present study, may be needed in order to stabilize the turbulence. This observation may have some relevance for the observed difference in the dynamics of electron and ion transport barriers in tokamak plasmas. More work, including a treatment of the coupled ITG/TE mode system using more realistic models, is needed in order to confirm this. This is left for future work.

\end{document}